\pdfoutput=1

\documentclass[11pt]{article}

\usepackage{naacl2021}
\usepackage{times}
\usepackage{booktabs}
\usepackage{latexsym}
\usepackage{multirow}
\usepackage{bbm}
\usepackage{algorithm}
\usepackage{algorithmic}
\usepackage{amsmath}
\usepackage{color}
\usepackage{graphicx}
\usepackage{subfigure}
\usepackage{enumitem}
\usepackage{comment}
\usepackage[normalem]{ulem}
\useunder{\uline}{\ul}{}

\usepackage[T1]{fontenc}

\usepackage[utf8]{inputenc}

\usepackage{microtype}

\newcommand{\ibbrfull}{In-Batch Balancing Regularization}
\newcommand{\ibbr}{IBBR}

\title{Debiasing Neural Retrieval via In-batch Balancing Regularization}

\author{Yuantong Li$^{1}$\Thanks{ Work done during an internship at AWS.}, Xiaokai Wei$^{2\dagger}$, Zijian Wang$^{2\dagger}$, Shen Wang$^{2}$\Thanks{ Equal contribution.}, \\ \textbf{Parminder Bhatia$^{2}$, Xiaofei Ma$^{2}$, Andrew Arnold$^{2}$} \\ $^1$UCLA \\  $^2$AWS AI Labs\\ \texttt{yuantongli@ucla.edu; {xiaokaiw, zijwan, shenwa, parmib}},\\ \texttt{xiaofeim, anarnld@amazon.com}}

\begin{document}

\maketitle
\begin{abstract}
People frequently interact with information retrieval (IR) systems, however, IR models exhibit biases and discrimination towards various demographics. 
The in-processing fair ranking methods provide
a trade-offs between accuracy and fairness through adding a fairness-related regularization term in the loss function. 
However, there haven't been intuitive objective functions that depend on the click probability and user engagement to directly optimize towards this.
In this work, 
we propose the \textbf{I}n-\textbf{B}atch \textbf{B}alancing \textbf{R}egularization (\ibbr{}) to mitigate the ranking disparity among subgroups. 
In particular, we develop a differentiable \textit{normed Pairwise Ranking Fairness} (nPRF) and leverage the T-statistics on top of nPRF over subgroups as a regularization to improve fairness.
Empirical results with the BERT-based neural rankers on the MS MARCO Passage Retrieval dataset with the human-annotated non-gendered queries benchmark \citep{rekabsaz2020neural} show that our \ibbr{} method with nPRF achieves significantly less bias with minimal degradation in ranking performance compared with the baseline.

\end{abstract}

\section{Introduction}
Recent advancements in Natural Language Processing and Information Retrieval \citep{palangi2016deep,devlin2018bert,zhao2020interactive,karpukhin2020dense} have led to great progress in search performances. However, search engines easily expose various biases (e.g.,  \citep{biega2018equity,baeza2018bias,rekabsaz2020neural,rekabsaz2021societal}), which sabotage the trust of human beings from day to day.
Many methods have been proposed recently to reduce the bias of the retrievers. Existing fairness-aware ranking methods can be categorized into pre-processing methods, in-processing methods, and post-processing methods \citep{mehrabi2021survey,zehlike2021fairness}. 
Pre-processing methods typically focus on mitigating bias in data before training the model. 
\citet{lahoti2019ifair} discussed the individual fairness pre-processing method to learn the fair representation of data. However, the representation-based method will undermine the value of the features determined by domain experts \citep{zehlike2021fairness}. The in-processing methods usually transform the fairness in ranking task into an optimization problem consisting of an accuracy objective and a fairness objective. These methods learn the best balance between these two objectives \citep{kamishima2011fairness, berk2017convex, bellamy2018ai, konstantinov2021fairness}.
\citet{zehlike2020reducing} handles different types of bias without knowing the exact bias form; 
Post-processing algorithms \citep{singh2018fairness,zehlike2017fa,zehlike2020matching,cui2021towards} are model agnostic without requiring access to the training process,  but these methods re-order the ranking at the expense of accuracy \citep{menon2018cost}. 

Among recent works on fair neural retrieval, \citet{beutel2019fairness} introduce the pairwise ranking fairness (PRF) metric for ranking predictions. This pairwise fairness metric evaluates whether there is a difference in accuracy between two groups.
\citet{rekabsaz2021societal} (AdvBert) mitigates the bias magnitude from the concatenation of query and passage text rather than treating the bias magnitude from query and passage separately through an adversarial neural network.

In this paper, we propose the In-Batch Balancing Regularization (IBBR) method combined with the neural retrieval model. IBBR is an in-processing debiasing method that balances the ranking disparity among different demographic groups by adding an in-batch balancing regularization term to the objective function. We design two batch-level regularization terms, \textit{Pairwise Difference} (PD) and \textit{T-statistics} (TS) that measure biases within demographic groups. In addition, we introduce normed Pairwise Ranking Fairness (nPRF), a relaxed version of the PRF \citep{beutel2019fairness} that is differentiable, thus could be directly optimized. We apply IBBR to MS MARCO passage re-ranking task \citep{nguyen2016ms} on gender bias using pre-trained $\text{BERT}_{L_{2}}$ and $\text{BERT}_{L_{4}}$ models \citep{turc2019well}. Empirical results show that our model could achieve significantly less bias with minor ranking performance degradation, striking a good balance between accuracy and fairness. 
Our contributions can be summarized as follows:
\begin{itemize}[leftmargin=*]
   \item We introduce \ibbr{}, an in-processing debiasing method based on pairwise difference and T-statistics. 
    \item We introduce normed PRF, a relaxed version of the pairwise ranking fairness (PRF) metric \citep{beutel2019fairness}. The normed PRF solves the non-differentiable issue and could be directly optimized during training. 
    \item We perform experiments on the MS MARCO passage re-ranking task with \ibbr{} and normed PRF. Empirical results show that \ibbr{} and normed PRF could achieve a statistically significant improvement in fairness while maintaining good ranking performance.
\end{itemize}

\begin{figure}[t]
   \centering
   \includegraphics[height=0.45\textwidth, width=0.47\textwidth]{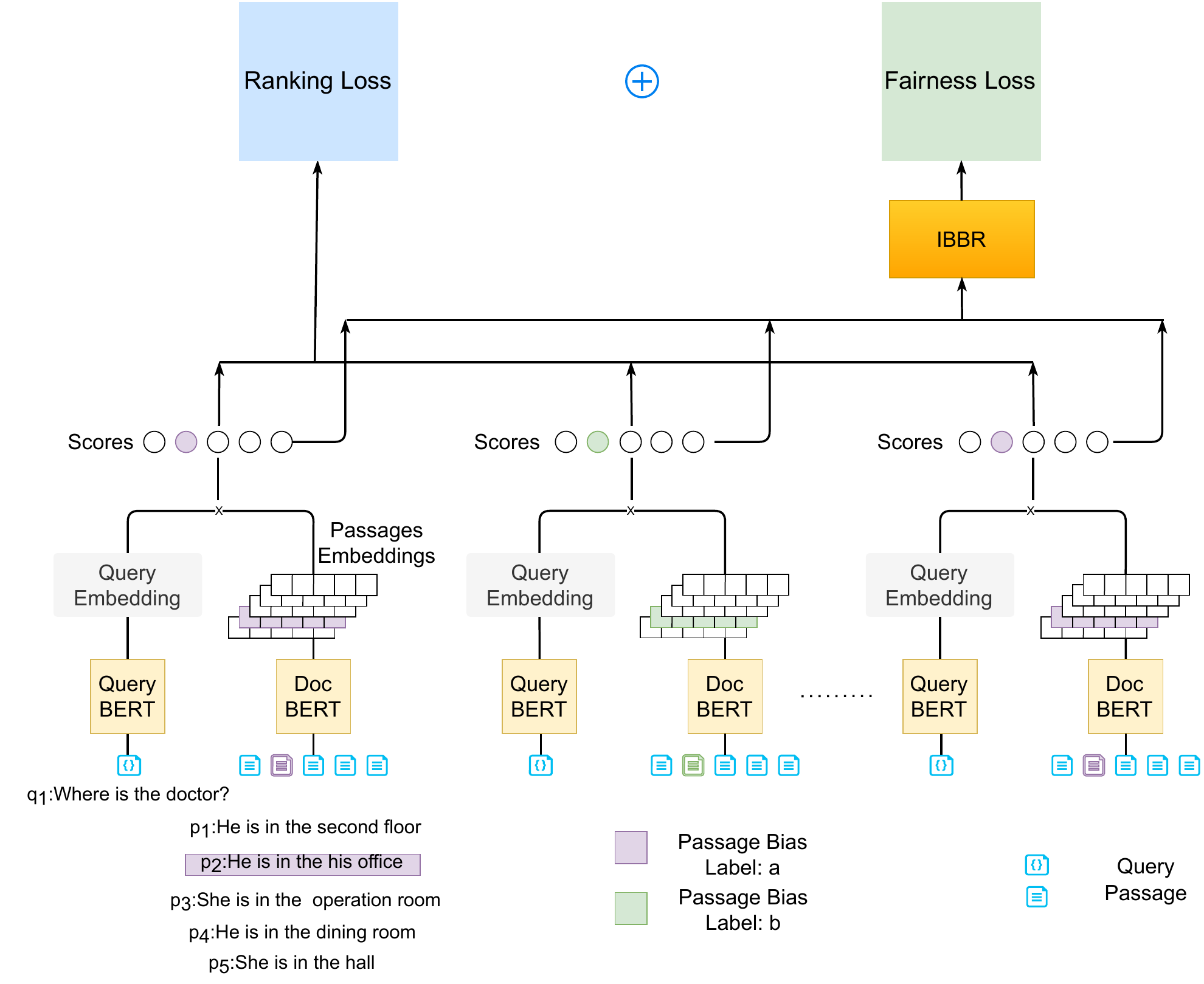}
   \caption{An example of \ibbrfull{} method. For each query, we calculate the typical ranking loss and the fairness loss from \ibbr{} on top $K$ retrieved passages. We jointly optimize the ranking loss and the fairness loss.
   There are two ways of computing the \ibbr{}, pairwise difference loss and T-statistics Loss. 
   }
   \label{fig:fig1}
   \vspace{-5mm}
\end{figure}

\section{PROBLEM DEFINITION}
\label{Sec: Task}
We first introduce notations in the ranking task in \S \ref{sec: Notation}. \S \ref{Bias level of a passage} provides the definition of the bias of the passage. In \S \ref{sec: group fair in ranking}, we propose the definition of the group fairness in the ranking task.

\subsection{Notations in the Ranking Task}
\label{sec: Notation}
Formally, we define the task of \textit{Gender Debiased Neural Retrieval} (GDNR) as: given a query $q$ and top $K$ passages retrieved by the neural retrieval system, we adapt the ranking to mitigate bias in the retrieval result.
We first define the whole query set as $Q = \{q_{1}, q_{2}, ..., q_{N}\}$. For each query $q_{i}$, we denote $P_{i} = \{p_{i,1}, p_{i,2}, \ldots, p_{i,j}, \ldots, p_{i,K}\}$ as the corresponding retrieved passages' set for query $q_{i}$. 
With query $q_{i}$ and corresponding retrieved passages $P_{i}$, $s_{i} = \{q_{i}, p_{i,1}^{+}, p_{i,2}^{-}, \ldots, p_{i, K}^{-}\}$ is defined as one data pair.
Here $p_{i,1}^{+}$ is the ground truth passage (clicked passage) and $p_{i,j}^{-}$ is the non-clicked passage, $\forall j \in \{2,3,...,K\}$. 
We use $Y_{i,j} = 1$ to label the passage $p_{j}$ as a clicked passage, otherwise, $Y_{i,j} = 0$. 
Finally, the whole dataset is defined as $D = \{s_{1}, s_{2}, ..., s_{N}\}$. 
For notation simplicity,
we use $[1:K]$ to represent $\{1,2,..., K\}$.

\subsection{Bias Label of Passage}
\label{Bias level of a passage}
We first provide the definition of the bias label of one passage, and consider the gender bias as a running example. \citet{rekabsaz2020neural} use the degree of gender magnitude  in the passage to define the bias value, where the gender concept is defined via using a set of highly representative gender definitional words. 
Such a gender definitional set usually contains words such as \textit{she, woman, grandma} for female definitional words ($G_{f}$), and \textit{he, man, grandpa} for males definitional words ($G_{m}$). 

The definition of the bias of the passage in our method is different from \citep{rekabsaz2020neural} who assume that one passage has two magnitudes: female magnitude and male magnitude. 
However, we assume that one passage has only one implication or tendency and use the gender magnitude difference as the bias value. 
So the bias value of the passage $p$, $mag(p)$, defined as
\begin{equation}
    mag(p) = \sum_{w \in G_{m}} \log |\langle w, p\rangle| - \sum_{w \in G_{f}} \log |\langle w, p\rangle|,
\end{equation}
where $|\langle w, p \rangle|$ refers to the number of occurrences of the word $w$ in passage $p$, $w \in G_{m}$ or $G_{f}$. 
Furthermore, we define the bias label for the passage $p$ as $d(p)$, if $mag(p) > 0$, then $d(p) = 1$ (male-biased); if $mag(p) < 0$, then $d(p) = -1$ (female-biased); if $mag(p) = 0$, then $d(p) = 0$ (neutral).
So for each retrieved passage $p_{i,j}, j \in [1:K], i \in [1:N]$, it has one corresponding bias label $d(p_{i,j}) \in \{-1, 0, 1\}$. 

\subsection{Group Fairness in Ranking}
\label{sec: group fair in ranking}
In \S \ref{sec: PRF}, 
we introduce one metric of ranking group fairness (pairwise ranking fairness) proposed by \citep{beutel2019fairness}.
In \S \ref{sec: nPRF}, we provide a more refined definition of pairwise ranking fairness.

\subsubsection{Pairwise Ranking Fairness}
\label{sec: PRF}
If $R(p)\in [0,1]$ is the ranking score of passage $p$ from one retrieval model , $\text{PRF}_{m}(s_{i})$ measures the probability level of a male-biased random passage selected from the male group $m$ higher than all random female-biased passages of data pair $s_{i}$
\begin{equation}
    \begin{aligned}
       \text{PRF}_{m}& (s_{i}) =  \frac{1}{n_{1}^{m}(s_{i})n_{0}(s_{i})}
        \sum_{j \in g_{1}^{m}(s_{i})}\\
        & \sum_{k \in g_{0}(s_{i})} \mathbbm{1}[R(p_{i,j}) \geq R(p_{i,k})],
    \end{aligned}
    \vspace{-2mm}
\end{equation}
where $g_{1}^{m}(s_{i}) = \{j|d(p_{i,j})=1, Y_{i,j} = 1,  j \in [1:K]\}$ represents passages clicked ($Y_{i,j} = 1$) as well as belonging to male biased group ($d(p_{i,j})=1$). $n_{1}^{m}(s_{i}) = |g_{1}^{m}(s_{i})|$ represents the number of male-biased clicked  passages. $g_{0}(s_{i}) = \{j|Y_{i,j} = 0, j \in [1:K]\}$ represents the group of non-clicked passages. $n_{0}(s_{i})$ represents the number of all non-clicked samples in retrieved passages. 
\citet{beutel2019fairness} use the probability that a clicked sample is ranked above another  non-clicked sample for the sample query as the pairwise accuracy. The pairwise fairness asks whether there is a difference between two groups when considering the pairwise accuracy as the fairness level metric.

However, we find that PRF is not directly applicable as an argument in the regularizer of a loss function  that works as a trade-off of accuracy and fairness. Because PRF is a \textit{0-normed objective function}, which is non-convex and non-differentiable. So we propose a modified PRF that can be optimized directly.

\subsubsection{Normed-Pairwise Ranking Fairness}
\label{sec: nPRF}
We propose a relaxed version of PRF called normed-PRF (nPRF), which measures the degree of group fairness in retrieval results for a given query and considering the ranking performance as well. The detailed definition of $\text{nPRF}_{m}$ is defined over all clicked male-biased passages $p_{i}$ in a data pair $s$ is
\begin{equation}
    \begin{aligned}
          &\text{nPRF}_{m}= \biggl\{ \frac{1}{n_{1}^{m}(s_{i})n_{0}(s_{i})}  \sum_{j \in g_{1}^{m}(s_{i})}  \\
         &\sum_{k \in g_{0}(s_{i})} |R(p_{i,j})|^{2} \mathbbm{1}[R(p_{i,j}) \geq R(p_{i,k})]\biggl\}^{\frac{1}{2}},
    \end{aligned}
\end{equation}
where $n_{1}^{m}(s_{i})$ is the number of all clicked male-biased passages in a data pair, usually $n_{1}^{m}(s_{i}) = 1$ in the ranking system.

In order to avoid the drawback of PRF being non-differentiable, 
we multiply the square of the ranking score ($|R(p_{i,j})|^{2}$) of the passage $p_{j}$ to the indicator function $\mathbbm{1}[R(p_{i,j}) \geq R(p_{i,k})]$, which is differentiable.  
Besides, $\frac{1}{n_{0}(s_{i})}\sum_{j \in g_{0}(s_{i})}|R(p_{i,j})|^{2}\mathbbm{1}[R(p_{i,j}) \geq R(p_{j,k})]$ measures the average harm of the biased passage $p_{i,j}$. If this value is large, it means that on average, these non-clicked passages are more relevant to the clicked passage $p_{i,j}$. This contributes more harm to the society since people are more willing to accept the ranking result.
If this value is small, it means that on average, these non-clicked passages are less irrelevant to the click passage $p_{i}$. This contributes less harm to the society since people are less willing to accept the ranking result.
Thus, the nPRF not only considers the magnitude of the ranking performance of the retrieval results but also inherits the explainable society impact into the PRF.

\section{Algorithms}
\label{sec: 5}
In this section, we create a regularizer based on the nPRF to mitigate the gender bias.
In \S \ref{sec: rank model}, we introduce necessary components for the neural retrieval task. 
In \S \ref{sec: loss function}, we provide the definition of the ranking loss and two fairness loss functions, \textit{Pairwise Difference Loss} and \textit{T-statistics Loss}, acting as a regularizer, named as \textit{in-batch balancing regularization method} (IBBR).

\subsection{Rank Model}
\label{sec: rank model}
Given the data set $D$,
we use the two-tower dense passage retriever (DPR) model \citep{karpukhin2020dense} as our retrieval model.  DPR uses two dense encoders $E_{P}, E_{Q}$ which map the given text passage and input query to two $d$-dimensional vectors ($d = 128$) and retrieves $K$ of these vectors which are close to the query vector. 
We define the ranking score between the query and the passage using the dot product of their vectors produced from DPR as $\text{sim}(q_{i},p_{i,j}) = z_{q_{i}}^\top z_{p_{i ,j}}$, where $z_{q_{i}} = E_{Q}(q_{i})$ and $z_{p_{i ,j}} = E_{P}(p_{i, j})$ are the corresponding query and passage dense embeddings.

\paragraph{Remarks.} 
Here we use two-tower DPR for two reasons. (I)   Computational considerations.  
\citet{humeau2019poly} thoroughly discussed the pros and cons between cross-encoders \citep{nogueira2019passage} and bi-encoders such as DPR and stated that cross-encoders are too slow for practical use. 
(II) Using cross-encoders can cause ill-defined problem such as, 
if the query's bias label belongs to groups $m$ and the passage's bias label belongs to group $f$, the concatenation of these two texts' bias label is unclear, based on the definition provided in Eq. (2) from \citep{rekabsaz2021societal}. 
So the two-tower BERT model is applied separately on the query and document to tackle this ill-defined problem.
Here we only consider the DPR as our ranking model.

\paragraph{Encoders.} 
In our work, in order to demonstrate the robustness of IBBR, we use two BERT models \citep{turc2019well}, (1) tiny BERT (base, uncased); (2) mini BERT (base, uncased) as our encoders, and take the representation at the [CLS] token as the output.

\paragraph{Inference.} For the data pair $s_{i}$, the ranking score $R(p_{i,j})$ of passage $p_{j}$ for query $q_{i}$ is simply the inner product of $\text{sim}(q_{i},p_{i,j})$ produced by DPR encoders.

\subsection{Loss Functions}
\label{sec: loss function}

\subsubsection{Ranking Loss}
The ranking loss is the \textit{negative log-likelihood loss} by computing the inner product of query and passage embeddings to measure the ranking performance for the data pair $s_{i}$,
\begin{equation*}
    \begin{aligned}
        \text{L}^{\text{Rank}} = -\log \frac{e^{(\text{sim}(q_{i}, p_{i,1}^{+}))}}{ e^{(\text{sim}(q_{i}, p_{i,1}^{+}))} +
        \sum_{j=2}^{K}e^{(\text{sim}(q_{i}, p_{i,j}^{-}))}}.
    \end{aligned}
\end{equation*}

\subsubsection{Fairness Loss}
To mitigate the bias for two groups, 
we use the ranking disparity as a measure to evaluate the fairness level of the neural retrieval system. 
And this ranking disparity works as a regularization in the loss function.
Here we propose two regularization terms as follows.

\paragraph{(I) Pairwise Difference Loss.} The pairwise difference (PD) loss $\text{L}^{\text{Fair}}_{\text{P}}$ measures the average ranking disparity between two groups $m$ and $f$ over a batch size ($B$) of data pairs,
\begin{equation}
    \begin{aligned}
        \text{L}^{\text{Fair}}_{\text{P}} &= \frac{1}{n_{m}n_{f}}
        \sum_{c \in p_{[1:B]}^{m}} 
        \sum_{d \in p_{[1:B]}^{f}} \\
        &(\text{nPRF}_{m}(s_{c}) - \text{nPRF}_{m}(s_{d}))^2,
    \end{aligned}
\end{equation}
where $P_{[1:B]}^{m} = \{i| p_{i,j} \in g^{m}_{1}(s_{i}), i \in [1:B], j \in  [1:K]\}$ is the set that the clicked passage belongs to group $m$ over batch size $B$ data. $P_{[1:B]}^{f} = \{i| p_{i,j} \in g^{f}_{1}(s_{i}),  i \in [1:B], j \in  [1:K]\}$ is the set that the clicked passage belongs to group $f$ over batch size $B$ data,
and $n_{m} = |P_{[1:B]}^{m}|$ and $n_{f} = |P_{[1:B]}^{f}|$. 

\paragraph{Remarks.} If there are many $\text{nPRF}_{m}(s_{x})$ which are different from other $\text{nPRF}_{f}(s_{y})$, this means that group $m$ and group $f$ have different fairness level over this batch data and will introduce more loss. 
However, this PD loss does not consider distribution information over this batch data, and imbalanced-data issue when group $m$ and group $f$ samples are imbalanced.
Thus we propose the T-statistics loss to overcome this.

\paragraph{(II) T-statistics Loss.}
The design of T-statistics (TS) loss is also based on the ranking disparity but considers the second order information (variance effect) of each group for each batch data. We use the square of T-statistics as the ranking disparity measure and defined as,
\begin{equation*}
    \begin{aligned}
        \text{L}^{\text{Fair}}_{T} = \big\{(\hat{\mu}_{m} - \hat{\mu}_{f})^{2}/\sqrt{\hat{\text{var}}_{m}/n_m +\hat{\text{var}}_{f}/n_f}\big\}^2,
    \end{aligned}
\end{equation*}
where $\hat{\mu}_{m} = \frac{1}{n_{m}}\sum_{j \in P_{[1:B]}^{m}} \text{nPRF}_{m}(j)$ is the mean of the male group's nPRF, and $\hat{\text{var}}_{m}=\frac{1}{n_{m}}\sum_{j \in P_{[1:B]}^{m}} (\text{nPRF}_{m}(j) - \hat{\mu}_{m})^2$ is the variance of the male group's nPRF.  Besides, $\hat{\mu}_{f}, \hat{\text{var}}_{f}$ can be defined similarly.

\paragraph{Remarks.}
This TS loss can provide a robust measure for the ranking disparity especially when the batch data pair is imbalanced. 
The square of the T-statistics, i.e., $\chi^{2}$ distribution, provides the theoretical guarantee and power to reject the similarity between group $m$ and group $f$. 

\paragraph{Total Loss.}
The total loss will be the sum of the ranking loss and fairness loss, represented as $L^{\text{total}}_{[1:B]} = L^{\text{rank}}_{[1:B]} + \lambda L^{\text{fair}}_{[1:B]}$,
where $L^{\text{fair}}$ can be the PD loss or TS Loss. $\lambda$ is a hyperparameter to control the balance of the fairness loss and ranking loss. In the experiment, we try manually and automatically to tune $\lambda_{\text{fair}}$. The details of our method can be found in Figure \ref{fig:fig1}.

\section{Experiments}
\label{sec: 6}
In this section, we describe data resources in \S \ref{sec: data description}, experiment setup in \S \ref{sec: exp setup}, evaluation metrics in Section \ref{sec: eva metrics}, baseline models in Section \ref{sec: baseline model}, and corresponding result analysis in Section \ref{sec: result analysis}.

\begin{table*}[]
\scriptsize
\centering
\begin{tabular}{ccccccccc}
\toprule
\multicolumn{3}{c}{\textbf{nPRF}}                  & \multicolumn{3}{c}{\textbf{Ranking Metric}}                                   & \multicolumn{3}{c}{\textbf{Fairness Metric}}                                  \\ \hline
\multirow{7}{*}{}       & \multicolumn{1}{c}{\textbf{IBBR}} & \multicolumn{1}{c}{$\lambda_{\text{fair}}$} & \textbf{Recall@10} $\uparrow$  & \textbf{MRR} $\uparrow$ & \textbf{NDCG} $\uparrow$ & \textbf{$|\Delta\text{A-PRF}|$} $\downarrow$ & \textbf{RaB@5} $\downarrow$  & \textbf{RaB@10} $\downarrow$             \\ \hline
\multirow{5}{*}{DPR BERT(L2)}      &   & & 0.357 & 0.164 & 0.196 & 0.005  & 0.091 & 0.079  \\ \cline{2-9}
    
 & \multirow{2}{*}{\textbf{PD}}     & Best   & 0.238 (-33.3\%)  & 0.112 (-31.7\%) & 0.124 (-36.7\%) & 0.034 (+580\%) & 0.094 (+3.3\%) & 0.083 (+5.1\%) \\  
     &  & Auto  & 0.270 (-24.3\%) & 0.126 (-23.2\%) & 0.143 (-27.0\%) & 0.033 (+560\%)  & 0.098 (+7.7\%) & 0.083 (+5.1\%)                 \\ %
& \multirow{2}{*}{\textbf{TS}} & Best  & \textbf{0.349 (-2.2\%)}  & \textbf{0.170 (+3.6\%)}  & \textbf{0.198 (+1.0\%)} & \textbf{0.001}$^{\ddag}$ \textbf{(-80\%)} & \textbf{0.091 (0\%)}   & \textbf{0.075}$^{\ddag}$ \textbf{(-5.1\%)}\\ 
&  & Auto  & 0.333 (-6.7\%)  & 0.160 (-2.4\%)   & 0.185 (-5.6\%)  & 0.006 (+20\%) & 0.109 (+19.7\%) & 0.077 (-2.5\%) \\ \midrule
\multirow{5}{*}{DPR BERT(L4)}   & & & 0.429 & 0.205 & 0.243 & 0.043  & 0.016  & 0.011 \\ \cline{2-9}

 & \multirow{2}{*}{\textbf{PD}} & Best & 0.381 (-11.1\%) & 0.213 (+3.9\%) & \textbf{0.236 (-2.8\%)}  & 0.034$^{\ddag}$ (-20.9\%) & 0.033 (+106\%) & 0.025 (+127\%) \\ 
 &  & Auto  & 0.373 (13.1\%) & \textbf{0.214 (+4.4\%)} & 0.234 (-3.7\%) & 0.030$^\ddag$ (-30.2\%) & 0.033 (+106\%) & 0.021 (+90.9\%) \\ %
 & \multirow{2}{*}{\textbf{TS}} & Best & 0.365 (-14.9\%) & 0.193 (-5.9\%) & 0.217 (-10.7\%) & \textbf{0.000}$^{\ddag}$ \textbf{(-100\%)} & \textbf{0.003}$^{\ddag}$ \textbf{(-81.3\%)} & \textbf{0.012} \textbf{(+9.1\%)}  \\ 
&  & Auto  & \textbf{0.389 (-9.3\%)}  & 0.205 (0\%) & 0.234 (-3.7\%)  & 0.022$^{\ddag}$ (-48.8\%) & 0.004$^{\ddag}$ \textbf{(-75.0\%)} & 0.017 (+54.5\%)\\ \bottomrule
\end{tabular}
\caption{
The ranking and fairness results of two IBBR methods, pairwise difference and T-statistics, combined with nPRF in $\text{BERT}_{L_{2}}$ and $\text{BERT}_{L_{4}}$ models. We compare IBBR with baseline models DPR L2, L4 in the re-ranking tasks and experimenting with different fairness hyperparameter $\lambda_{\text{fair}}$ tuning methods.
The bold value in each column shows the best result in that metric.
$\uparrow$ and $\downarrow$ indicate larger/smaller is better in corresponding definition of metrics. 
$^{\ddag}$ indicates statistically significant improvement (p-value$<0.05$) over the DPR baseline in fairness metrics. 
}
\label{table: 2-PRF}
\vspace{-1mm}
\end{table*}
\subsection{Dataset}
\label{sec: data description}
We experimented on the passages re-ranking task from MS MARCO \citep{nguyen2016ms}. This collection includes 8.8 million passages and 0.5 million queries, comprised of  question-style queries from Bing's search logs, accompanied by human-annotated clicked/non-clicked passages. Additionally, data bias labels over this dataset are available from \citep{rekabsaz2020neural}.

\paragraph{Data For DPR.} The whole dataset is composed of total 537,585 queries and $K * 537,585$ retrieved passages where $K = 200$, for the baseline DPR model. Each query has top $K$ passages including one ground truth and 199 negative samples.
The details of splitting the dataset used for training, development, and test (7:2:1) for the DPR model can be found in Appendix \ref{sec:appendix} Table \ref{table: data for DPR}. There are 126 queries used for the final evaluation. 

\paragraph{Data For Fair Model.} The fairness dataset \citep{rekabsaz2020neural} is also created upon this MS MARCO dataset. These queries were annotated into one of four categories: non-gendered (1765), female (742), male (1,202), other or multiple genders (41).
Here we only use the non-gendered queries, and assume the query is unbiased given it does not have any gender definitional terms. 
There are 1,252 unique queries in total. Examples of non-gendered queries are: \textit{what is a synonym for beautiful?},  \textit{what is the meaning of resurrect?}, etc.

\vspace{-3pt}

\subsection{Experiment Setup}
\label{sec: exp setup}
The maximum length of query and passage are set to 100. Batch size $B$ is 150 optimized over $\{100, 120, 150\}$. Learning rate is $3e^{-5}$ optimized over $\{3e^{-6},3e^{-5}, 3e^{-4}\}$. 
A warmup ratio of 10\% with linear scheduler and a weight decay of 0.01 are set. %
In addition, we searched the fairness penalty parameter $\lambda = [0.1, 0.5, 1, 5, 10]$ (Best). We also experimented setting the $\lambda_{\text{fair}}$ as a trainable parameter (Auto). All experiments are conducted ten times and we reported the average.

\subsection{Evaluation Metrics}
\label{sec: eva metrics}
\paragraph{Ranking metrics.}
We use Recall@10, MRR, and NDCG@10 to evaluate the ranking performance. 

\paragraph{Fairness metrics.}
We use RaB@5, RaB@10, and ranking disparity $|\Delta\text{A-PRF}|$ to evaluate the fairness magnitude.

\paragraph{$\text{RaB}_{t}$.}  $\text{RaB}_{t}$ is a measurement of ranking bias, which is based on the average of the gender magnitude of passages at top $t$ ranking list \citep{rekabsaz2020neural}. To measure the retrieval bias, RaB calculates the mean of the gender magnitudes of the top $t$ (5 or 10) retrieved documents for the data pair $s_{i}$, for females,
$\text{qRaB}_{t}^{f}(s_{i}) = \frac{1}{t}\sum_{j=1}^{t} mag_{f}(p_{i,j})$.
Using these values, the RaB metric of the query $q$, $\text{RaB}_{t}(s_{i})= \text{qRaB}_{t}^{m}(s_{i}) -  \text{qRaB}_{t}^{f}(s_{i})$, and the RaB metric of the retrieval model over all the queries,
$\text{RaB}_{t} = \frac{1}{N}\sum_{s_{i}\in D}\text{RaB}_{t}(s_{i})$.
The smaller the absolute value of $\text{RaB}_{t}$, the less the ranking disparity is.

\paragraph{$|\Delta\text{A-PRF}|$.}
$|\Delta\text{A-PRF}|$ measures the ranking disparity over two groups, which is the difference over two averaged PRF,
$|\Delta\text{A-PRF}| = |\frac{1}{|T_{m}|} \sum_{i \in T_{m}} \text{PRF}_{i} - \frac{1}{|T_{f}|} \sum_{i \in T_{f}} \text{PRF}_{i}|$, where $T_{m}$ is the dataset that the clicked passage belongs to group $m$, $T_{m} = \{i| Y_{i,j} = 1, d_{i,j} = 1, \forall i \in [1:N]\}$. With the running example, we denote $|T_{m}|$ as the number of male-biased clicked pairs and similar definitions are for $T_{f}$ and $|T_{f}|$. 
The smaller the $|\Delta\text{A-PRF}|$ is, the smaller the ranking disparity is. If $|\Delta\text{A-PRF}|$ is close to zero, it means that the retrieved results are relatively fair since the two groups' PRF are close to each other. 
To avoid selection bias, $|\Delta\text{A-PRF}|$ measures the whole dataset's fairness level rather than the subset's result such as top 5 and top 10. 
\vspace{-2mm}

\subsection{Baseline Models}
\label{sec: baseline model}
 The baseline methods contain the classical IR models, BM25, and RM3 PRF, and neural based models: Match Pyramid (MP), Kernel-based Neural Ranking Model (KNRM), Convolutional KNRM (C-KNRM), Transformer-Kernel (TK), and the fine-tuned BERT Model. These results are available in in Appendix Section \ref{sec:appendix}. For the BERT rankers, we use BERT-Tiny ($\text{BERT}_{L_{2}}$) and BERT-Mini ($\text{BERT}_{L_{4}}$).

\subsection{Results Analysis}
\label{sec: result analysis}
\paragraph{Ranking Performance.} In Table \ref{table: 2-PRF}, we present the result of original $\text{BERT}_{L_{2}}$ and $\text{BERT}_{L_{4}}$ and  $\text{BERT}_{L_{2}}$ and $\text{BERT}_{L_{4}}$ with IBBR (PD and TS). 
We found that in $\text{BERT}_{L_{2}}$, after adding IBBR, the ranking performance decreases 2.2\% in Recall@10 and the bias level decreases 80\% when applying the TS. Overall, TS outperforms PD on average when considering the ranking metrics because it downgrades the ranking metric less, which can be found in the ranking metric columns. This phenomenon exists both in hand-tuned or auto-tuned hyperparameter $\lambda_{\text{fair}}$ and $\text{BERT}_{L_{2}}$ and $\text{BERT}_{L_{4}}$. 

\paragraph{Fairness Performance $|\Delta\text{A-PRF}|$.}  
$\text{BERT}_{L_{2}}$ + TS can achieve \textit{$80\%$} reduction in mitigating $|\Delta\text{A-PRF}|$ bias.
The $|\Delta\text{A-PRF}|$ fairness metric in $\text{BERT}_{L_{4}}$+TS can achieve \textit{$100\%$} reduction in mitigating bias compared with the original $\text{BERT}_{L_{4}}$.
Besides, PD performs unsatisfied in the fairness metric compared with TS in $\text{BERT}_{L_{2}}$ and $\text{BERT}_{L_{4}}$, we found that the variance of nPRF and the imbalance affects the performance of PD, which is usually found in the training phase ($\#$male-biased > $\#$female-biased). Overall nPRF + TS can achieve the best performance in mitigating the $|\Delta\text{A-PRF}|$ ranking disparity, which achieves our goal in mitigating the ranking disparity.

\paragraph{Fairness Performance RaB.} As for RaB, we hope to use another fairness metric to demonstrate our regularization's robustness. 
We realize RaB is focusing on the top-ranking result and $|\Delta\text{A-PRF}|$ is focusing on the overall ranking result by definition. We present the RaB result in the last column.
In the last two columns, the TS method is still better than the PD method on average. 
For RaB@5, the TS method's performance is similar to the PD method in $\text{BERT}_{L_{2}}$ (3.3\% vs 0\%);
The TS method's performance is better than the PD method in $\text{BERT}_{L_4}$ (106\% vs -81.3\%).
For RaB@10, in $\text{BERT}_{L_{2}}$, the TS method is similar to the PD method (5.1\% vs -2.5\%);
In $\text{BERT}_{L_{4}}$, the TS method is better than the PD method (90.9\% vs 9.1\%).
After evaluating the the fairness level on $\text{BERT}_{L_{4}}$ and $\text{BERT}_{L_{2}}$, we found that the more complicated the model is, the more bias it is, which is also demonstrated in \citep{rekabsaz2020neural}. 
We find that the RaB performance not consistent with the $|\Delta\text{A-PRF}|$ is mainly because $|\Delta\text{A-PRF}|$ is focusing more on the lower-ranked passages and RaB is focusing the higher-ranked passages. 
This makes these two fairness metrics are relatively exclusive. However, when the ranking system performs well (rank the clicked passage high), the $|\Delta\text{A-PRF}|$  will finally consider the overall ranking result.

\section{Conclusion}
In this paper, we present a novel in-processing in-batch balancing regularization method to mitigate ranking disparity and retain ranking performance. We also overcome the non-differentiable and non-convex properties of the 0-normed PRF and propose the nPRF. We conduct experiments on the MS MARCO dataset and find that the nPRF with T-statistics regularization method outperforms other methods in terms of fairness metrics and ranking metrics. 
In future work, we will consider generalizing our method to multiple protected variables such as age, income, etc, and also addressing bias in the query by employing adversarial networks.

\section*{Bias Statement}
In this paper, we study gender bias in the neural retrieval system. If a ranking system allocates resources or opportunities unfairly to specific gender groups (e.g., less favorable to females), this creates allocation harm by exhibiting more and more male-dominated passages, which also forms a more biased dataset in turn. When such a ranking system is used in reality, there is an additional risk of unequal performance across genders. Our work is to explore the bias level of the dense passage retrieval model with $\text{BERT}_{L_{2}}$ and $\text{BERT}_{L_{4}}$ on the MS MARCO passage reranking task. Thus, the community can use these benchmarks with a clearer understanding of the bias level, and can work towards developing a fairer model.

\bibliographystyle{acl_natbib}
\bibliography{anthology}

\appendix

\section{Appendix}
\label{sec:appendix}
\begin{table*}[!htbp]
\centering
\begin{tabular}{cccccccc}
\toprule
Model  &  \multicolumn{1}{c}{} & \multicolumn{3}{c}{\textbf{Ranking Metric}} & \multicolumn{3}{c}{\textbf{Fairness Metric}}  \\ \hline
\multirow{7}{*}{}       &  & \textbf{Recall@10}  & \textbf{MRR} & \textbf{NDCG} & \textbf{D-PRF} & \textbf{RaB@5}  & \textbf{RaB@10}              \\ \hline
    \textbf{BM25}   &   & 0.230 & 0.107 & 0.125 & - & - & -  \\
    \textbf{RM3 PRF}     & & 0.209 & 0.085 & 0.104 & - & - & - \\ 
    \textbf{MP}   &   & 0.295 & 0.141 & 0.191 & - & - & - \\ 
    \textbf{KNRM}     & & 0.297 & 0.169 & 0.167 & - & - & - \\ 
   \textbf{C-KNRM}      & & 0.325 & 0.170 & 0.197 & - & - & - \\ 
    \textbf{TK}   &   & 0.360 & 0.212 & 0.231 & - & - & - \\ \cline{1-8}
    \textbf{DPR(L2)}     & & 0.357$^{\dag}$ & 0.164$^{\dag}$ & 0.196$^{\dag}$ & 0.005  & 0.091 & 0.079  \\
    \textbf{DPR(L4)}    & & 0.429$^{\dag}$ & 0.205$^{\dag}$  & 0.243$^{\dag}$ & -0.043 & 0.016 & 0.011  \\
\bottomrule
\end{tabular}
\caption{IR Model and DPR, $^{\dag}$ indicates significant improvement over BM25. 
}
\label{table: IR model}
\end{table*}
In this section, we provide the baseline model performance in Table \ref{table: IR model}. We also provide the training, development, and test of the origin dataset and the fairness dataset (with fairness label) in Table \ref{table: data for DPR}.

\begin{table*}[!htbp]
\centering
\begin{tabular}{lrrr|r}
\toprule
\multicolumn{1}{c}{\textbf{Data}}  &
\multicolumn{1}{c}{\textbf{Train}} &  \multicolumn{1}{c}{\textbf{Dev}} & 
\multicolumn{1}{c}{\textbf{Test}} & 
\multicolumn{1}{|c}{\textbf{Total}} \\ \hline
    \multirow{1}{*}{DPR} & 510,586  & 26,873 & 126 & 537,585 \\ \cline{2-5}
    \multirow{1}{*}{Fairness\footnote{Here we use Non-gendered queries}} & 876  & 250  &  126 & 1,252\\  \bottomrule
\end{tabular}
\caption{The number of training, development and testing examples for the DPR model and fairness model}
\label{table: data for DPR}
\end{table*}

\end{document}